\documentstyle{article}
\begin{document}
\topmargin -1cm 

\vspace{1cm}

\centerline{\bf Distributions of Time- and Distance-Headways in the}
\centerline{\bf Nagel-Schreckenberg Model of Vehicular Traffic:}
\centerline{\bf Effects of Hindrances}
\vspace{1cm}
\centerline{\bf Debashish Chowdhury$^1$, Abhay Pasupathy and Shishir Sinha}
\centerline{Physics Department, Indian Institute of Technology} 
\centerline{Kanpur 208016, India}
\vspace{1cm}
\begin{abstract}
In the Nagel-Schreckenberg model of vehicular traffic on single-lane highways 
vehicles are modelled as particles which hop forward from one site to another 
on a one dimensional lattice and the inter-particle interactions mimic the 
manner in which the real vehicles influence each other's motion. In this model 
the number of empty lattice sites in front of a particle is taken to be a 
measure of the corresponding {\it distance-headway}({\bf DH}). The {\it 
time-headway}({\bf TH}) is defined as the time interval between the 
departures (or arrivals) of two successive particles recorded by a detector 
placed at a fixed position on the model highway. We investigate the effects 
of spatial inhomogeneities of the highway (static hindrances) on the DH and 
TH distributions in the steady-state of this model. 

\end{abstract}

\vspace{1cm}




\vspace{1cm}

\noindent-----------------------------------------------------------------------------------------

$^1$To whom all correspondence should be addressed; E-mail: debch@iitk.ernet.in

\vfill\eject 
\renewcommand{\baselinestretch}{2}\huge\normalsize

\section{Introduction:}

The continuum models of traffic flow [1-3]  are analogues of the 
"hydrodynamic" models of fluid flow while the kinetic theories of 
vehicular traffic [4-6], which are extensions of the kinetic 
theory of gases, and the car-following models [7-10] as well 
as the discrete "particle-hopping" models [11-17] are analogues 
of the "microscopic" models of interacting particles commonly studied in 
statistical mechanics. In this paper we focus our attention on a 
specific particle-hopping model, namely, the {\it Nagel-Schreckenberg} 
({\bf NS}) model [11] of vehicular traffic on idealized single-lane 
highways; this model may be regarded as a model of interacting 
particles driven far from equilibrium and the dynamical phenomena 
exhibited by this model of traffic may be treated as problems of 
non-equilibrium statistical mechanics. 

The {\it distance-headway}({\bf DH}) is defined as the distance from a 
selected point on the lead vehicle ({\bf LV}) to the same point on the 
following vehicle ({\bf FV}). Usually, the front edges or bumpers are 
selected [18]. The {\it time-headway}({\bf TH}) is defined as the time 
interval between the departures (or arrivals) of two successive vehicles 
recorded by a detector placed at a fixed position on the highway  [18]. 
The DH and TH are not merely of academic interest to statistical 
physicists, but are also of practical interest in traffic engineering 
where these two distributions are regarded as important characteristic 
of traffic flow [18,19]. For example, larger headways provide greater 
margins of safety whereas higher capacities of the highway require smaller 
headways. 

Recent investigations by several groups of statistical physicists [20-24] 
have helped in gaining insight into the nature of the "jamming transition" 
(i.e., the transition from the "free-flowing" dynamical phase to the 
"jammed" dynamical phase) in the NS model. On the other hand, simultaneously, 
NS model is being generalized and extended [25-37] to capture more and 
more details of vehicular traffic flow so that the generalized/extended 
model may, ultimately, find practical use in traffic engineering [38-39]. 
Extending our earlier numerical works on the DH distribution in the 
steady-state of the NS model, we point out here interesting similarities 
as well as crucial differences between the "jamming transition" and the 
gas-liquid phase transition in a simple fluid in equilibrium. Moreover, 
in this paper we also present a simple analytical calculation of the TH 
distribution in the steady-state of a special case of the NS model using 
an interesting quantity introduced in ref.[15]. Furthermore, we study the 
effects of spatial inhomogeneities of the highway (static hindrances) on 
the DH and TH distributions in the steady-state of the NS model.

The NS model and some of its most important features are summarized in 
section 2. We report numerical results on the DH distribution in the steady-
state of the NS model, in the absence as well as in the presence of 
hindrances, in section 3 where we also draw attention to the analogies 
and differences between the jamming transition in the NS model and the 
gas-liquid transition. Our calculations of the TH distribution in the 
steady-state of the NS model are given in the section 4, where we also 
study the effects of hindrances on this distribution. Finally, we 
summarize our main results and conclusions in section 5. 

\section{The Models:} 

In the NS model a lane is represented by a one-dimensional 
lattice of $L$ sites. Each of the lattice sites can be either empty or 
occupied by at most one "vehicle". If periodic boundary condition is 
imposed, the density $c$ of the vehicles is $N/L$ where $N (\leq L)$ is 
the total number of vehicles. In the NS model [11] the speed $V$ of each 
vehicle can take one of the $V_{max}+1$ allowed {\it integer} values 
$V=0,1,...,V_{max}$. Suppose, $V_n$ is the speed of the $n$-th vehicle 
at time $t$. At each {\it discrete time} step $t \rightarrow t+1$, the 
arrangement of $N$ vehicles is updated {\it in parallel} according to 
the following "rules":

\noindent {\it Step 1: Acceleration.} If, $ V_n < V_{max}$, the speed 
of the $n$-th vehicle is increased by one, i.e., $V_n \rightarrow V_n+1$.

\noindent{\it Step 2: Deceleration (due to other vehicles).} If $d_n$ is the 
gap in between the $n$-th vehicle and the vehicle in front of it, and if 
$d_n \le V_n$, the speed of the $n$-th vehicle is reduced to $d_n-1$, i.e., 
$V_n \rightarrow d_n-1$.

\noindent{\it Step 3: Randomization.} If $V_n > 0$, the speed of the $n$-th 
vehicle is decreased randomly by unity (i.e., $V_n \rightarrow V_n-1$) with 
probability $p$ ($0 \leq p \leq 1$); $p$, the random deceleration probability, 
is identical for all the vehicles and does not change during the updating.

\noindent{\it Step 4: Vehicle movement.} Each vehicle is moved forward so 
that $X_n \rightarrow  X_n + V_n$ where $X_n$ denotes the position of the 
$n$-th vehicle at time $t$. 

The specific update rule of the NS model requires a nonvanishing braking 
probability $p$ for the model to yield a realistic description of traffic 
flow [12] and, thus, the NS model may be regarded as stochastic cellular 
automata [40]. Effectively free flow of traffic takes place when the 
density of vehicles is sufficiently low whereas high density leads to 
congestion and traffic jams. One of our aims is to point out some 
similarities as well as differences between the "jamming transition" 
in the NS model [20-24], which is a non-equilibrium driven system, and 
the gas-liquid phase transition in a fluid in equilibrium. 

The asymmetric simple exclusion process (ASEP) is one of the simplest 
models of driven systems of interacting particles; it is often treated 
as a carricature of traffic flow. The relations between this model and 
the $V_{max} = 1$ limit of the NS model has been elucidated in the 
literature [24]. Some dramatic effects of quenched disorder (static 
hindrances) on the steady state of this model have been investigated 
[41,42]. Our main aim in this paper is to investigate the effects of 
static hindrances on the DH and TH distributions in the NS model.

Emmerich and Rank [32]  introduced an extra step of update rule {\it before} 
all the other steps of updating in the NS model to mimic the effects of static  
traffic {\it hindrance}: a segment of length $L_{hind}$ on the highway is 
identified as the hindrance and the speed of all the vehicles found 
within that segment of the highway are reduced to {\it half} of their current 
speed. A more general model of traffic flow in the presence of "quenched 
disorder" in the highway was formulated by Csahok and Vicsek [33] in 
terms of an "inverse permeability" associated with each site. We shall use 
the simpler rule introduced by Emmerich and Rank [32]. 

For the convenience of our analytical calculations, following Schreckenberg 
et al.[14], we assume the sequence of steps $2-3-4-1$, instead of $1-2-3-4$; 
the advantage is that there is no vehicle with $V = 0$ immediately after the 
acceleration step. Consequently, if $V_{max}=1$, we can then use a binary 
site variable $\sigma$ to describe the state of each site; $\sigma=0$ 
represents an empty site and $\sigma = 1$ represents a site occupied by a 
vehicle whose speed is unity. 

\section{Distance-Headway Distributions and the Nature of the Jamming Transition} 
Using the sequence of steps $2-3-4-1$, as explained earlier, an $n$-cluster 
configuration in the steady-state of the NS model is represented by 
$(\sigma_1, \sigma_2,...,\sigma_n)$. The number of empty lattice sites, 
$n$, in front of a vehicle is taken to be a measure of the corresponding DH.  
Within the 2-cluster approximation, the DH distribution, 
${\cal P}_{2c}^{dh}(n)$, 
in the steady-state of the NS model with $V_{max} = 1$ has been calculated 
analytically following the methods of the {\it site-oriented} mean-field 
({\bf SOMF}) theory [14]. In this approximation,  
\begin{equation}
{\cal P}_{2c}^{dh}(0) = {\cal C}(\underline{1}|1)
\end{equation}
and
\begin{equation}
{\cal P}_{2c}^{dh}(n) = {\cal C}(\underline{1}|0)\left\{ {\cal C}(\underline{0}|0)
\right\}^{n-1}{\cal C}(\underline{0}|1), \quad for \quad n \geq 1
\end{equation}
where, ${\cal C}$ gives the 2-cluster steady-state configurational 
probability for the argument configuration and the underlined imply
the conditional, as usual. The expressions for the various ${\cal C}$s
are given by [14,24] 
\begin{equation}
{\cal C}(\underline{0}|0) \quad =\quad {\cal C}(0|\underline{0}) = 1 - \frac{y}{d} 
\end{equation}
\begin{equation}
{\cal C}(\underline{1}|0) \quad =\quad {\cal C}(0|\underline{1}) = \frac{y}{c} 
\end{equation}
\begin{equation}
{\cal C}(\underline{0}|1) \quad =\quad {\cal C}(1|\underline{0}) =  \frac{y}{d} 
\end{equation}
\begin{equation}
{\cal C}(\underline{1}|1) \quad =\quad {\cal C}(1|\underline{1}) = 1 - \frac{y}{c} 
\end{equation}
where 
\begin{equation}
y = \frac{1}{2q}\left( 1 - \sqrt{1 - 4 q c d}\right), 
\end{equation} 
$q = 1 - p$ and $d = 1 - c$. 
It turned out that the 2-cluster approxmation is exact for $V_{max} = 1$ 
[14]. Therefore, in terms of $c$ and $p$, the exact DH distribution in the 
steady-state of the NS model with $V_{max} = 1$ is given by        
\begin{equation}
{\cal P}^{dh}(0) = 1 - (y/c) 
\end{equation}
and 
\begin{equation}
{\cal P}^{dh}(n) = \{y^2/(cd)\}[1 - (y/d)]^{n-1} \quad for \quad n \geq 1
\end{equation} 
where, for the given $c$ and $p$, $y$ can be obtained from equation (7). 
The same exact DH distribution (8-9) has also been derived 
independently [15] within the framework of {\it car-oriented} mean-field 
({\bf COMF}) theory. The DH distributions in the NS model for all 
$V_{max} > 1$ have been computed numerically by carrying out computer 
simulation [24]. 

The DH distribution in the steady-state of the NS model simultaneously 
exhibits two peaks over an intermediate regime of the vehicle densities,  
provided $V_{max} >1$ and $p$ is sufficiently large. In contrast, when 
$V_{max} = 1$, no such two-peak structure is exhibited at any density, for 
any $p$, by the DH distribution which is given by the exact expression (8-9).
We have now estimated, as functions of $p$, the densities, $c_s$ and 
$c_{\ell}$, corresponding to the smallest and the largest densities where 
two-peak structure is exhibited by the DH distribution in the NS model 
($2 \leq V_{max} \leq 5$).  $c_s$ and $c_{\ell}$ are plotted against $1-p$ 
for $V_{max} = 2$ in fig.1. 

The occurrence of the two-peak structure for $V_{max} > 1$ has been interpreted 
[24,43] as a signature of "two-phase coexistence"; the two coexisting phases, 
namely, the "free-flowing phase" and the "jammed phase" being the analogues of 
the gas phase and the liquid phase, respectively, of a simple fluid in 
equilibrium. In this scenario, for a given $p$ (and given $V_{max}$), it is 
tempting to identify $c_s$ as the analogue of the density corresponding to 
the onset of two-phase coexistence in a fluid while $c_{\ell}$ would be the 
corresponding largest density up to which these two phases continue to coexist 
at a given temperature; accordingly, the regime of density in between these 
two curves would be the analogue of the two-phase coexistence region for a 
fluid and the uppermost tip of this region could be interpreted as an analogue 
of the critical point. Further, this analogy would suggest that $1-p$ is the 
analogue of temperature; {\it increase} of $p$ (i.e., decrease of $1-p$) 
drives the system towards "condensation". 

In spite of this apparent analogy, there are crucial differences between 
the "jamming transition" and the gas-liquid transition in a simple fluid. 
First of all, the fluid is in equilibrium whereas the model vehicular traffic 
under consideration is in a non-equlibrium steady-state. More important 
difference is that the size of the jams remain finite even in the limit of 
infinite system size [24] and, therefore, cannot be identified as a true 
dynamical phase. 

When the DH distribution exhibits two peaks simultaneously, the peak at 
vanishing DH reflects the fact that the {\it total} number of vehicles held 
up simultaneously in the various jams on this model highway is a finite 
fraction of all the vehicles on this highway; these jams arise from {\it 
spontaneous} fluctuations in the driven system of interacting particles and, 
therefore, can appear anywhere in the system. In contrast, a hindrance can 
induce recurring jams at (and near) its own location, as demonstrated 
in fig.2, because of the bottlenecks against traffic flow created by it. 
The DH distributions for 5 different values of $L_{hind}$ are plotted 
in fig.3, all for the same density $c = 0.05$ and for the same $p = 0.5$. 
In this figure the absence of two-peak structure for $L_{hind} = 0$ indicates 
that, for $p = 0.5$, the density $c = 0.05$ is not high enough to have a 
finite fraction of the vehicles, simultaneously, held up in various jams 
created by spontaneous fluctuations. However, for the same set of values 
of $c$ and $p$, two-peak structure occurs in the DH distribution with the 
increase of $L_{hind}$; this is a consequence of jams caused by the bottleneck 
effect of the hindrance. 

\section{Time-Headway Distribution:}

There are several earlier papers, published by statisticians and traffic 
engineers, where the form of the TH distribution has been derived on the 
basis of heuristic arguments [44]. In this section we begin by presenting 
a simple derivation of this distribution for $V_{max} = 1$ using a quantity 
introduced in ref.[15] in the context of the COMF theory. However, we could 
compute the corresponding TH distributions for $V_{max} > 1$ [45] and the 
effects of hindrances on these distributions only through computer simulation. 

We label the position of the detector by $j=0$, the site immediately in 
front of it by $j=1$, and so on. The detector clock resets to $t=0$ everytime 
a vehicle leaves the detector site. 
We begin our analytical calculations for $V_{max} = 1$ by writing 
${\cal P}^{th}(t)$, the probability of a time headway $t$ between a LV 
and the FV, as 
\begin{equation}
{\cal P}^{th}(t) = \sum_{t_1=1}^{t-1} P(t_1)Q'(t-t_1|t_1) 
\end{equation}
where $P(t_1)$ is the probability that there is a time interval $t_1$ 
between the departure of the LV and the arrival of the FV at the detector 
site and $Q'(t-t_1|t_1)$ is the conditional probability that the FV halts 
for $t - t_1$ time steps when it arrived at the detector site $t_1$ time 
steps after the departure of the LV. 

Suppose, $g(t)$ is the probability that a vehicle moves in the next 
(i.e., in the $t+1$-th time step). Therefore, $\bar{g}(t) = 1 - g(t)$ 
is the probability that a vehicle does not move in the next time step. 
It has been shown [15] that $g = q[1 - {\cal P}^{dh}(0)]$. Using equation 
(8) for ${\cal P}^{dh}(0)$ we get 
\begin{equation}
g = qy/c  
\end{equation}
and, hence, $\bar{g} = 1-qy/c$, where $y$ is given by the equation (7).  
Moreover, since $y$ satisfies the equation $qy^2 - y + cd $ = 0, it 
follows that
\begin{equation}
(1-qy/c)(1-qy/d) = p.
\end{equation}

In order to calculate $P(t_1)$ we need to consider all those spatial 
configurations at $t=0$ from which the FV can reach the detector site 
within $t_1$ steps. For all configurations with $t_1>n$, $t_1-1$ time 
steps elapse in crossing $n - 1$ linkss (as the last link is crossed 
certainly at the last time step). Thus 
\begin{equation}
P(t_1) = \sum_{n=1}^{t_1} \Pi(n) q^n p^{t_1-n} \quad ^{t_1-1}C_{n-1} 
\end{equation}
where
\begin{equation}
\Pi(n) = {\cal C}(1|\underline{0})\left\{ {\cal C}(0|\underline{0}) \right\}^{n-1}.  
\end{equation}
and, hence, we get 
\begin{equation}
P(t_1)  =  {\cal C}(1|\underline{0})q
	\left[{\cal C}(0|\underline{0})q + p \right]^{t_1-1} 
  = \frac{qy}{d}(1-\frac{qy}{d})^{t_1-1}  
\end{equation}
which is the exact analytical expression for $P(t_1)$.

Next, we calculate $Q(t-t_1\mid t_1)$ by expressing it in terms of $g$  
and $\bar{g}$. When the FV arrives at the detector site exactly $t_1$ 
time steps after the departure of the LV, the LV can be at any of the 
sites labeled by $1,\cdots,t_1+1$. The probability that the LV stays at 
the site '1' is $\bar{g}^{t_1}$ and, therefore, the probability that 
the LV is not at the site '1' is $1 - \bar{g}^{t_1}$. 

If the LV is not at site '1' then the probability that the FV halts at 
the detector site for exactly $t-t_1$ time steps is $p^{t-t_1-1}q$ because 
it should stop due to randomisation for exactly $t-t_1-1$ steps and move at 
the last step. Hence, the contribution to $Q'(t-t_1\mid t_1)$ when the LV 
is not at site '1' is
$$(1 - \bar{g}^{t_1})p^{t-t_1-1}q \eqno(I) $$ 
On the other hand, when the LV is at site '1' then it will have to move so 
that the FV is able to leave the detector site after $t-t_1$ steps. Suppose 
the LV moves from '1' after $k$ steps ($k$ varies from 1 to $t-t_1-1$). Then 
the FV will have to stay at the detector site for next $t-t_1-1-k$ steps 
due to randomisation and move in the last step. Hence, the contribution to 
$Q'(t-t_1\mid t_1)$ when LV is at site '1' is
$$ \bar{g}^{t_1}gq( \sum_{k=1}^{t-t_1-1}\bar{g}^{k-1}p^{t-t_1-1-k})
= \bar{g}^{t_1}gq \frac{[(\bar{g})^{t-t_1-1} - (p)^{t-t_1-1}]}{\bar{g}-p} \eqno(II) $$
Therefore, combining (I) and (II) we get 
\begin{equation}
Q'(t-t_1\mid t_1) = (1 - \bar{g}^{t_1})p^{t-t_1-1}q + 
 \bar{g}^{t_1}gq \frac{[(\bar{g})^{t-t_1-1} - (p)^{t-t_1-1}]}{\bar{g}-p} 
 \end{equation}
Finally, substituting the exact expressions (15) and (16) into (10) we get 
\begin{eqnarray} 
 {\cal P}^{th}(t) = \left[\frac{qy}{c-y}\right] \{1-(qy/c)\}^{t-1} + 
 \left[\frac{qy}{d-y}\right] \{1-(qy/d)\}^{t-1} \nonumber\\ 
 - \left[\frac{qy}{c-y}+\frac{qy}{d-y}\right] p^{t-1} - q^2(t-1)p^{t-2}.
\end{eqnarray}
where, for the given $c$ and $p$, $y$ can be obtained from equation (7). 
The invariance of the distribution (17) under the interchange of 
$c$ and $1-c$, is a manifestations of the well-known "particle-hole" 
symmetry in the problem which breaks down for all $V_{max} > 1$ [46]. 

The flux $q$ of the vehicles can be written as $q = N/T$ where 
$T = \sum_{i=1}^N t_i$ is the sum of the time headways recorded 
for all the $N$ vehicles. Therefore, one can rewrite $q$ as $q = 1/T_{av}$ 
where $T_{av} = (1/N)\sum_i t_i $ is the average TH.  Therefore, $T_{av}$ 
is expected to exhibit a minimum at $c = c_m$ with the variation of 
density $c$ of the vehicles. We observed that the trend of variation of 
the most-probable TH, $T_{mp}$, with $c$ is similar to that of $T_{av}$ 
with $c$ [45]. Moreover, because of the particle-hole symmetry, the $T_{mp}$ 
versus $c$ curve is symmetric about $c = 1/2$ in the NS model with 
$V_{max} = 1$. But this symmetry is lost when $V_{max} > 1$. The fact that 
$T_{mp}$ (and $T_{av}$) versus $c$ curve exhibits a minimum is consistent 
with one's intuitive expectation that both at very low and very high 
densities there are long time gaps in between the departures of two 
successive vehicles from a given site.  

Two typical sets of curves showing the effects of the hindrances on the 
TH distribution in the steady-state of the NS model are shown in fig.4. 
At low density of the vehicles, increase of the length of the hindrance 
leads to significant broadening of the distribution although its effect 
on the magnitute of the most probable TH is weak (see fig.4a). On the other 
hand, at moderate and high densities, the broadening of the TH distribution 
caused by increase of $L_{hind}$ is much weaker (see fig.4b). 

\section{Summary and Conclusion:} 

In this paper we have investigated two important characteristics of traffic 
flow on highways, namely, DH and TH distributions, starting from models 
that incorporate explicitly vehicle-vehicle and road-vehicle interactions. 
We have compared and contrasted the effects of {\it static hindrance} on the 
highway and those of the LV, which may be viewed as a {\it dynamic hindrance}, 
against the forward movement of the FV. In the absence of any static 
hindrance jams can appear anywhere in the system because of spontaneous 
fluctuations and the dynamic hindrance caused by a slower LV on the FV. 
On the other hand, in the presence of a static hindrance, jams take place 
at (and near) the hindrance, in addition to those formed by spontaneous 
fluctuations, because of bottleneck it creates against the traffic flow.

We have extracted some informations on the "structures" in the {\it spatial} 
oraganization of the vehicles on the highway from the DH distribution. 
We have demonstrated the effects of the bottleneck against traffic flow 
created by static hindrances on the DH distribution thereby elucidating the 
physical meaning of the two-peak structures observed earlier in the DH 
distribution over some intermediate regime of density when $V_{max} > 1$ 
and $p$ is sufficiently large. 

The TH distributions in the steady state of the NS model and the trend of 
their variation with density are in good {\it qualitative} agreement with 
the corresponding empirical data [18]. Our results demonstrate that, in 
spite of its simplicity, the NS model captures the essential qualitative 
features of the TH distribution of vehicular traffic on highways. Besides, 
the extent of broadening of the TH distribution caused by a hindrance depends 
on the density of the vehicles; the higher is the density the larger is 
the broadening. Hoever, at all densities, the effect of the hindrance on the 
magnitude of the $T_{mp}$ is very weak. 

In this paper we have considered only the original version of the NS 
model for vehicular traffic on single-lane highways [11] and an extended 
version that includes a specific type of inhomogeneities of the highway 
(namely, static hindrances) [32]. We have not attempted any direct {\it 
quantitative} comparison of our results with the corresponding empirical 
data from highway traffic because, we believe, such comparisons will be 
possible only after several of the realistic generalizations and extensions 
[25-37], proposed recently in the literature, are incorporated in the model. 
Results of our ongoing works in this direction will be published elsewhere 
[47].  

\vspace{.25cm}

{\bf Acknolwdegements:} We dedicate this paper to Professor Zittartz on 
the occassion of his 60th birthday. The research career of the senior 
author (DC) has been influenced strongly by the generous encouragements 
received from Professor Zittartz over the last 14 years. This paper is in 
an area where present Members of Professor Zittartz's group as well as his 
former students, postdocs and assistants have made most of the influential 
contributions. DC also thanks D. Stauffer, A. Schadschneider, L. Santen 
and anonymous referees for useful comments and the Alexander von Humboldt 
Foundation, for partial support through a research equipment grant.

\newpage
\noindent{\bf Figure Captions:} 

\noindent{\bf Fig.1}: $c_{\ell}$ and $c_s$ in the NS model are plotted 
against $1-p$ when $V_{max} = 2$ (see the text for the definitions of 
$c_{\ell}$ and $c_s$).

\noindent {\bf Fig.2}: The "space-time diagram" showing the time-evolution 
of the traffic in the NS model ($V_{max} = 5, p = 0.5$) in the presence of 
ten hindrances, each of length $5$, put randomly along the highway of length 
$1000$. Each of the black dots represents a vehicle. 

\noindent{\bf Fig.3}: The distance-headway distributions corresponding to 
the vehicle density $c = 0.05$ in the NS model with $V_{max} = 5$ 
($p = 0.5$) for five different values of the hindrance length are compared 
with that in the absence of hindrance. The discrete data points obtained 
from computer simulation correspond to  
$L_{hind} = 0 (+), L_{hind} = 1 (\times), L_{hind} = 2 (\ast), 
L_{hind} = 3 (\Box), L_{hind} = 4 (\rule{2mm}{2mm}), L_{hind} = 5 (\circ)$  
while the continuous curves are merely guides to the eye. 

\noindent {\bf Fig.4}: The time-headway distributions in the NS model 
($V_{max} = 5$), in the presence of hindrance, for vehicle densities 
(a) $c = 0.1$ and (b) $c = 0.5$. The symbols corresponding to 
$L_{hind} = 0, 1, 2, 3, 4$ and $5$ are identical to those in fig.3.

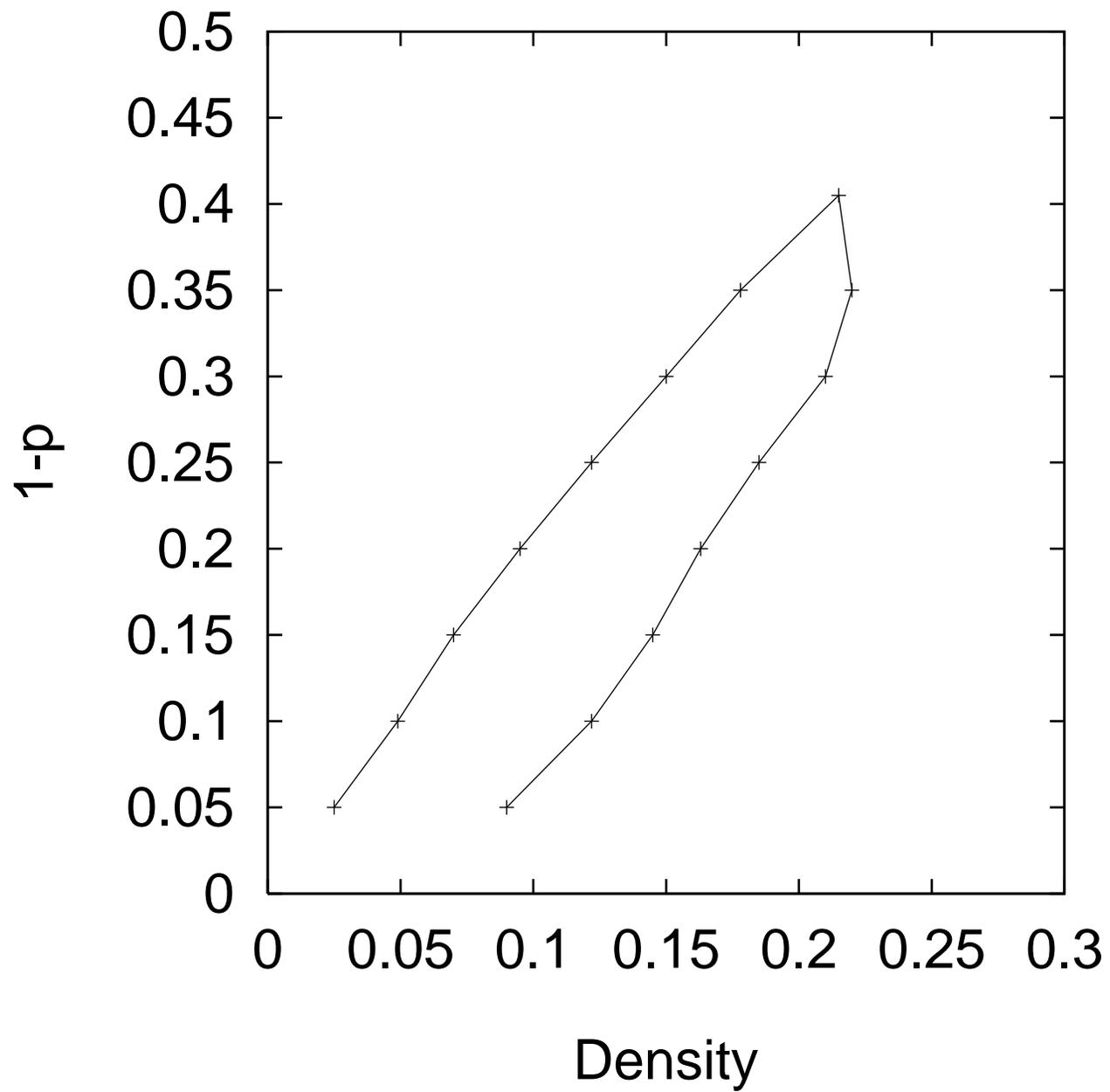
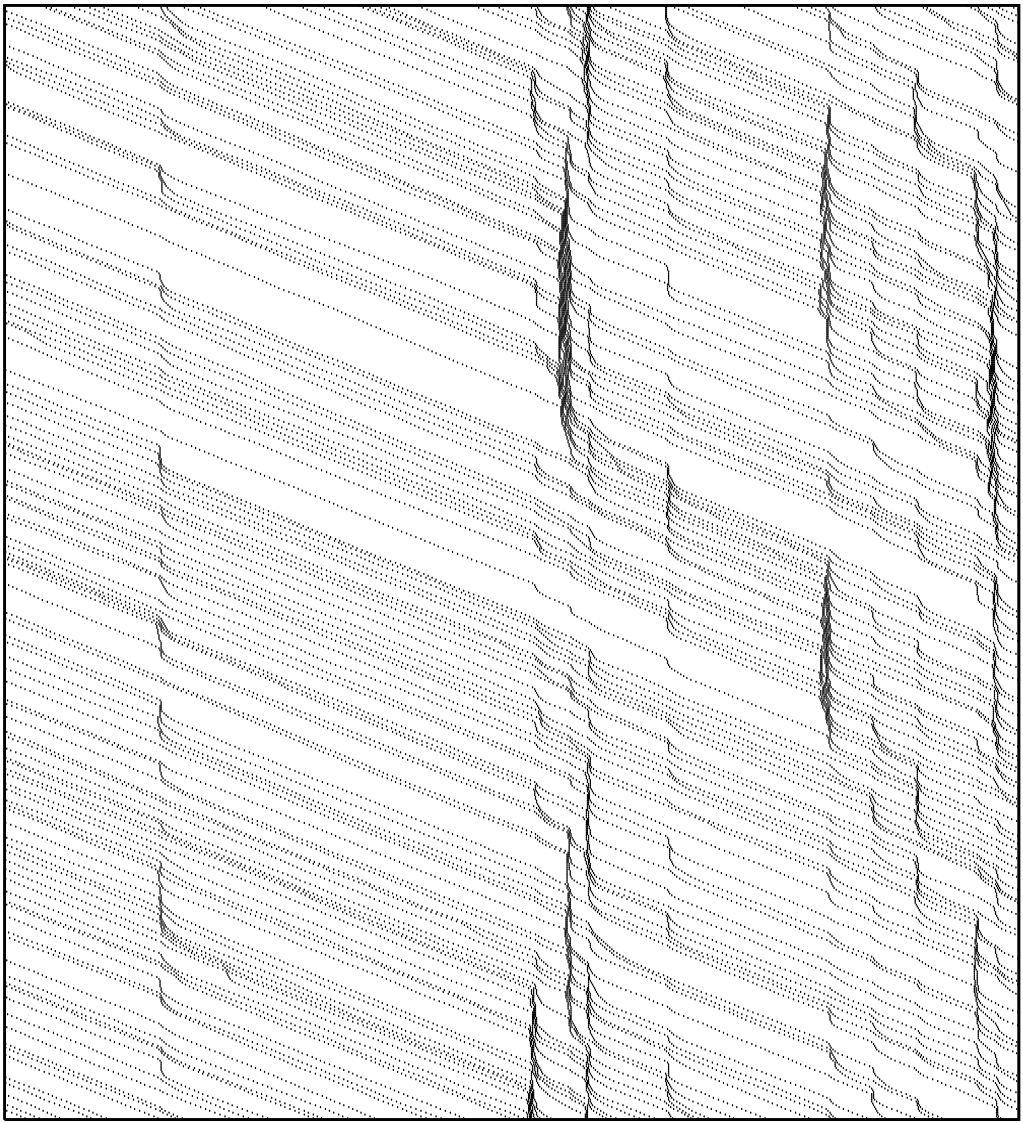
\end{document}